**A 3D joint interpretation of magnetotelluric and seismic tomographic models: the case of the volcanic island of Tenerife**

Araceli García-Yeguas[1,2,3], Juanjo Ledo[3,6], Perla Piña-Varas[6*], Janire Prudencio[4,2,3], Pilar Queralt[6], Alex Marcuello[6], Jesús M. Ibañez[2,5], Beatriz Benjumea[7], Alberto Sánchez-Alzola[8], Nemesio Pérez[3,9]

1. Department of Applied Physics. University of Cádiz. Avenida Universidad de Cádiz, 10, 11519 Puerto Real, Cádiz. Spain.

2. Instituto Andaluz de Geofísica. University of Granada. Granada. Spain.

3. Instituto Volcanológico de Canarias. INVOLCAN. Tenerife Island. Spain

4. Department of Computer Science. Georgia State University. 25 Park Place NE, Atlanta, GA 30302. USA

5. Department of Theoretical Physics and Cosmos. University of Granada. Granada. Spain.

6. GEOMODELS Research Institute, Departament de Dinàmica de la Terra i l'Oceà, Facultat de Geologia. Universitat de Barcelona. C/ Martí i Franquès s/n 08028 Barcelona, Spain

   *Now at Centre for Exploration Targeting, the University of Western Australia. 35 Stirling highway, Crawley, WA 6009

7. Institut Cartogràfic i Geològic de Catalunya. Parc de Montjüic s/n 08038 Barcelona. Spain.

8. Departamento de Estadística e Investigación Operativa, Universidad de Cádiz, Spain

9. Instituto Tecnológico de Energías Renovables. Pol. Ind. De Granadilla s/n. 38600. Granadilla de Abona. Santa Cruz de Tenerife (Spain).

**Abstract**

In this work we have done a 3D joint interpretation of magnetotelluric and seismic tomography models. Previously we have described different techniques to infer the inner structure of the Earth. We have focused on volcanic regions, specifically on Tenerife Island volcano (Canary Islands, Spain). In this area, magnetotelluric and seismic tomography studies have been done separately. The novelty of the present work is the combination of both techniques in Tenerife Island. For this aim we have applied Fuzzy Clusters Method at different depths obtaining several clusters or classes. From the results, a geothermal system has been inferred below Teide volcano, in the center of Tenerife Island. An edifice hydrothermally altered and full of fluids is situated below Teide, ending at 600 m below sea level. From this depth the resistivity and $V_P$ values increase downwards. We also observe a clay cap structure, a typical feature in geothermal systems related with low resistivity and low $V_P$ values.

Keywords: Seismic-tomography, magnetotelluric, Tenerife Island, cluster analysis.

# 1 Introduction

The volcanic island of Tenerife belongs to Canary Islands archipelago (Spain) (figure 1). Volcanism on Tenerife Island is very heterogeneous encompassing from basaltic eruptions to highly explosive eruptions (Romero 1991, 1992; Araña et al. 1994; Martí et al. 1994; Ablay et al. 1995; Dóniz et al. 2008; Andújar et al. 2008; Dóniz 2009). The most important edifice is located in the center of the island and is formed by Las Cañadas caldera and Teide-Pico Viejo stratovolcanoes Complex (CTPVC) in its center. Due to the gas exhalation from the active geothermal system beneath Teide, fumaroles can be observed often in its crater (Pérez et al. 1996). While the geomorphology and geology of the island are well known (Romero 1991, 1992; Martí et al. 1994; Dóniz et al. 2008; Dóniz, 2009) there are uncertainties related to its inner structure. For that reason, different geophysical studies have been carried out to determine the internal structure of Tenerife island: magnetotelluric (MT) measurements (Pous et al. 2002; Coppo et al. 2008; Piña-Varas et al. 2014, 2015); aeromagnetic surveying (Blanco-Montenegro et al. 2011), gravimetry studies (Araña et al. 2000; Gottsmann et al. 2008) or seismological studies (Canales et al. 2000; De Barros et al. 2012; García-Yeguas et al. 2012; Lodge et al. 2012; Prudencio et al. 2013; Prudencio et al. 2015), among others. However, there is a lack of attempts to jointly interpret the geophysical data and models acquired by the different agencies and institutions.

These techniques can be used separately, but could it be possible to combine them to obtain a better analysis of the inner Earth? The answer is affirmative. In this way the greatest advances in the knowledge of the internal structure and evolution of volcanoes are obtained from joint analysis and interpretation of different geophysical and geochemical observables (multi-method data sets). The multi-method data sets can contain information about the same property measured with different methods (i.e.

electrical conductivity measured with electrical resistivity tomography and MT methods) or about different properties (i.e. electrical conductivity and seismic velocity). In the first situation, the data sets are complementary and their combination will provide a more complete image of the process under observation (i.e. a more reliable and/or high resolution electrical conductivity model at different scales). In the second situation new information that would not be available from the single data sets (i.e. a lithological classification based on electrical conductivity and seismic velocity values). In all the cases, the main motivation for multi-sensor data fusion is to improve the quality of the output information combining the multi-sensor data inputs. A broad range of methodologies, going from a qualitative correlation to a quantitative joint inversion may be employed. A combination between qualitative and quantitative correlation methods based on clustering procedures, have been proposed by several authors (Bosch, 1999; Paasche and Tronicke, 2007; Bedrosian et al. 2007; Newmann et al. 2008; Stankiewicz et al. 2011; Falgàs et al. 2011; Shahrabi et al. 2015, among others). Statistical techniques have been recently applied in Dead Sea transform in Jordan (Bedrosian et al. 2007) or Groß Schönebeck geothermal site in Germany (Muñoz et al. 2010; Bauer et al. 2012) and fuzzy-logging methods have been applied recently to the Southwest of Iran (Shahrabi et al. 2015).

In the present work, we have carried out a joint interpretation of resistivity (Piña-Varas et al. 2014) and P-wave seismic velocity ($V_P$) models (García-Yeguas et al. 2012) using the Fuzzy Clusters Method (FCM) which its usefulness in the use of geophysical data have been demonstrated by Bezdek et al. (1984).

## 2 Data: Resistivity and seismic tomography models

## 2.1 Resistivity model

Piña-Varas et al. (2014, 2015) obtained a 3D resistivity model of Tenerife Island using a dataset of 188 broadband ($10^{-3}$-$10^2$ s) magnetotelluric soundings with the ModEm code (Egbert and Kelbert 2012). The model is discretised onto a 94x65x133-layer grid, and the inversions are undertaken using the off-diagonal components (Zxy,Zyx) of the impedance. In the inversion process, a 5 % error floor in the impedance components was imposed the final RMS is 2.3 after 50 iterations.

The MT model obtained by Piña-Varas et al. (2014, 2015) shows different areas according to their resistivity where the most important anomalies are the medium-high resistivity (100-500 Ω) body at the bottom of the model, the low resistivity area in the central region (<10 Ωm) and the region with low-medium values of resistivity (20-100 Ωm) surrounding the low resistivity area. It is important to remark that most of those structures are parallel to the topography. The authors interpreted these structures as different elements of the geothermal system: the body with medium-high resistivity correspond to the hottest part of the system, the structure with low-medium resistivity is related with rocks at higher temperatures and, finally, the clay cap overlying a geothermal reservoir is identified as the low resistivity layer. In order to asses the uncertainty of the MT model we have followed Piña-Varas et al. (2014) where they show that, as a first approximation, the electrical structure of the island can be simplified as a layered model, which follows the topography. Thus, assuming the 5% error used for the impedance tensor components during the inversion process, it can be show that the error in the 1D resistivity model is around 10% .

## 2.2  3D seismic velocity model

Garcia-Yeguas et al. (2012) obtained a 3D P-wave seismic velocity model using the dataset provided by an active seismic experiment (TOM-TEIDEVS) (Ibáñez et al.

2008). During the experiment 125 seismometers were deployed inland and more than 6300 shots were generated by air-guns fired by BIO Hespérides research vessel. In total, 103750 high quality travel times were selected to perform the tomographic inversion.

To check the robustness of the 3D seismic tomography model García-Yeguas et al. (2012) performed checkerboard, free anomalies and jackknifing tests. To calculate the P-wave velocity uncertainties associated to each depth of the model, these authors used the results of the checkerboard. They selected the central region on the island (axe x: 45 to 65km and axe y: 40 to 60km) of the recovered model and subtract the initial synthetic model. In this area there are eight anomalies of each type (low and high velocities). To the chosen values, they determinate the mean of the uncertainties with a confidence interval of 95%. The values for each depth are indicated in Table 1. To analyse the recuperated anomalies (low, high and medium anomalies), Garcia-Yeguas et al. (2012) made free anomalies tests, subtracting the recovered anomalies to the synthetic model. From these values they obtained the mean uncertainty for P-wave velocity for each type of anomaly. For a high velocity anomaly (15% of perturbation respect the initial model) the mean uncertainty is 0,2903 (km/s) in a 95% interval of confidence of [0,2674 y 0,3131] (km/s). In the case of a low velocity anomaly (-15% of perturbation) the mean uncertainty is higher, 0,5449 (km/s) in a 95% interval of confidence of [0,5164, 0,5734] (km/s). Considering a medium velocity anomaly (5% of perturbation) the mean is 0,1990 (km/s) in a 95% interval confidence of [0,1881, 0,2100] (km/s).

García-Yeguas et al. (2012) observed that a high $V_P$ core that characterizes the central structure of Tenerife Island reaching the surface in CTPVC. The authors interpreted it as the evidence of single central volcanic source during the formation of Tenerife Island. Low $V_P$ anomalies are mainly distributed around the high $V_P$ structure. They can be interpreted as fractured zones, hydrothermal alterations, porous materials and thick volcanoclastic deposits.

## 3    Fuzzy c-means clustering

The Fuzzy Cluster Method (FCM) or fuzzy c-means clustering divides the input dataset into $m$ fuzzy clusters, each of which is a fuzzy set in the sense that the boundaries between sets are poorly defined and possibly overlapped (Dunn 1973; Bezdek 1981). Hence, any data point may partially belong to several fuzzy clusters with different grades of membership. It has been applied in geophysics by several authors (Paasche and Tronicke 2007; Shahrabi et al. 2015 and references therein).

In the following we will explain the main procedures to apply FCM. Consider a finite set of elements ($X=\{x_1, x_2, \ldots., x_n\}$) as being p-dimensional, where p is the number of different datasets being analysed.

In this study, the elements will be the spatially coincident cells of the resistivity and velocity models forming a two-dimensional vector with the values of electrical resistivity and seismic velocity ($x_i=(\rho_i,v_i)$). The seismic model mesh grid had a 700x700x700 m discretization, the electrical resistivity model has a finer mesh grid. Thus, in order to generate a two-dimensional vector with the values of electrical resistivity and seismic velocity ($x_i=(\rho_i,v_i)$) the electrical resistivity model was downsampled into a 700x700x700 m mesh grid, resulting in a net loss of resistivity information but keeping the available seismic information.

The purpose of the method is to perform a partition of this set of elements (X) into c fuzzy sets (clusters) and at the same time minimize an objective function that measures the distance between the centre of each cluster and the data. The final result of fuzzy clustering can be expressed by a partition (membership) matrix U such that:

$$U=[u_{ij}]\ i=1\dots c,\ j=1\dots n$$

Where $u_{ij}$ is a numerical value in $[0,1]$ and expresses the degree to which the element $x_j$ belongs to the ith cluster. Two additional constraints in the $u_{ij}$ values are imposed, first a total membership of the element $x_j$ in all clusters is equal to 1, and second every constructed cluster is nonempty. In the fuzzy c-means algorithm developed by Bezdek (1981) and Bezdek et al. (1984) the objective function takes the form of

$$J\left(u_{ij}, \boldsymbol{v}_k\right) = \sum_{i=1}^{c}\sum_{j=1}^{n} u_{ij}^{m}\left\|\boldsymbol{x}_j - \boldsymbol{v}_i\right\|^{2}; m>1$$

where m is called the exponential weight which influences the degree of fuzziness of the membership matrix and $\boldsymbol{v}_i$ is the central vector of the ith cluster. The exponential, m, controls the relative weight placed on each of the squared differences. As m is closer to 1 the partitions that minimize the objective function become increasingly hard, in the sense that the element $x_j$ will belong only to one cluster ($u_{ij} \cong 1, u_{kj} \cong 0\ for\ k \neq i$). Increasing m tends to degrade (blur, defocus) membership towards the fuzziest state. No computational or theoretical evidence distinguishes and optimal value of m and for most of data $1.5 < m < 3$ gives good results (Bezdek et al., 1984). In our case a value of m=2 was selected, which is widely accepted as a suitable choice (Hathaway and Bezdek, 2001).

An important question for the fuzzy c-means algorithm is how to determine the correct number of clusters if no a priori information is available. Two parameters, the partitioning coefficient F and the partitioning entropy H (Burrough et al., 2000) computed as:

$$H(U,c) = \frac{1}{n}\sum_{j=1}^{n}\sum_{i=1}^{c}\left|u_{ij}ln u_{ij}\right|$$

$$F(U, c) = \frac{1}{n} \sum_{j=1}^{n} \sum_{i=1}^{c} u_{ij}^2$$

are defined to characterize each partition. A good classification has a combination of relatively large values of F and small values of H, although this fact has not been theoretically justified (Bezdek, 1981). In order to make F and H independent on the number of clusters (Guillaume, 2001), they can be scaled as:

$$F_s = \frac{F - \frac{1}{c}}{1 - \frac{1}{c}}$$

$$H_s = \frac{H - (1 - F)}{\ln(c) - (1 - F)}$$

Paascche et al. (2010) proposed another cluster validity measures but also point out that statically indicators could be complemented by other geophysical/geological data (i.e. well log information) to better determine the optimum number of clusters if they are available. The $F_s$ and $H_s$ coefficients are shown in figure 2 for different number of clusters. Following the criteria of maximizing $F_s$ and minimizing $H_s$, five clusters show the best results. Thus, not having a priori information, neither well-log values of resistivity and velocity, the final number of clusters chosen to carry out all the process will be five.

The fuzzy c-means geophysical model of Tenerife Island is estimated through the following procedure:

i) Conversion of the two independent 3D models into a common format, downsampling the electrical resistivity model of Piña-Varas et al. (2014). Thus each point on the new grid is associated with two values corresponding to the normalized electrical resistivity (log ρ) and the normalized seismic velocity (v).

ii) To solve the system we used an iterative fuzzy c-means approach provided in MATLAB (fcm function). As a result each (v,$log\rho$) pair is associated to a degree of membership to the different cluster centers.

iii) The defuzzification (Melgani et al., 2000) process consists of associate each (v,$log\rho$) point to the cluster center for which it has the highest degree of membership.

## 4    Results and discussion

We have obtained fuzzy c-means images for Tenerife Island at eight depths, every 700 meters from 2200 m (a.s.l.) to 2700 m (b.s.l.) using five clusters. Figure 3 shows the counting of (v,logr) values as well as the five cluster centers and the partitions of the cluster. This figure is segmented in several polygons. Each polygon represents the different clusters for both variables and the grey scale indicates the density of P-wave velocity and resistivity values.

Figures 4 and 5 show the results, a different colour is used for each cluster. The meaning of the colours are as follow. *Blue:* $\rho$ high and $V_P$ medium (RhVm);  *Light blue:* $\rho$ medium and $V_P$ low (RmVl); *Yellow:* $\rho$ low and $V_P$ low-medium (RlVlm); *Pink:* $\rho$ medium and $V_P$ high (RmVh); *Green:* $\rho$ medium and $V_P$ medium (RmV$_P$ m).   The values of the cluster centers  are the following, blue (RhVm): (5.47 km/s, 150 Ohm.m); light blue (RmVl): (3.76 km/s, 40 Ohm.m); yellow (RlVlm): (4.51 km/s, 6 Ohm.m); pink (RmVh): (5.97 km/s, 52 Ohm.m); green (RmVm): (5.13 km/s,  36 Ohm.m). The values of resistivity, velocity and uncertainties, which their possible geological interpretation have been described on Table 2.

Figure 4 shows two perpendicular cross-sections of the final 3D clustering image. The general distribution of the five clusters supports itself the use of all the data together (at different depths), in order to obtain a compact image that facilitates the interpretation of the results. We have taken into account the resolution aspects in our models to describe the images (see García-Yeguas et al., (2012) and Piña-Varas et al., (2014)).The most predominant feature is the high resistivity and medium $V_P$ cluster (blue colour) in the middle part of the figures. It increases its extension in latitude and longitude with depth. To the coast, the resistivity decreases, going from the high resistivity cluster to medium resistivity (pink and green colours) and low resistivity cluster (yellow colour). The $V_P$ shows a more complex pattern, slightly increasing from the center, blue colour, into a narrow zone, pink colour, and then decreasing progressively (green and yellow colours). The blue, pink and green center clusters all have a $V_P$ between 5-6 km/s that following Watts et al. 1997 will correspond to the central igneous core of the system. The $V_P$ differences between the blue (5.47 km/s) and pink (5.97 km/s) are relatively small, however the differentiation between these two clusters is imposed by the high difference of the resistivity centers, blue (150 Ohm.m) and pink (52 Ohm.m). In case of geothermal system the increase in $V_P$ (pink colour) could be attributed to the presence of medium to high temperature propylites with chlorite (Kanitpanyacharoen et al., 2011). This pattern of hydrothermal alteration represents the commonly observed in high-temperature geothermal systems (Ryan and Shalev, 2014; Frolova et al., 2014).

 On top of this three clusters lays the yellow one, which given its low resistivity (center cluster 6 Ohm.m, and all the values below 10 Ohm.m) may be linked   to the clay cap of the geothermal system or volcanoclastic deposits (surrounding CTPVC). The $V_P$ of the yellow cluster center (4.5 km/s) will be associated to a clay cap rich in smectite with low porosity (Tudge and Tobin, 2013). Figure 5 displays the obtained cluster distribution at different depths and in the following we discuss them arranged in four significant depth groups.

## 4.1 2200 m (a.s.l.) and 1500 m (a.s.l.)

At 2200 m (a.s.l.) (figure 5. C1 and C2) the whole area shows high $\rho$ and medium $V_P$, the east flank of Teide displays medium $\rho$ and low $V_P$ values (light blue colour). This region is called Montaña Blanca (Fig. 1 for location), the last plinian eruption at Tenerife, 2000 years ago. This structure is very fragmented and full of volcanic deposits (Ablay et al. 1995). At 1500 m (a.s.l.) CTPVC is characterized by medium-low $\rho$ and medium $V_P$ values (yellow and green colours) which can be related to the combined effect of shallow aquifers, sedimentary and volcanoclastic multifractured systems as interpreted in the 3D attenuation tomography by Prudencio et al. (2015). In the external areas of the complex higher values of both, $\rho$ and $V_P$, are obtained (blue and pink colours).

## 4.2 800 m (a.s.l.) and 100 m (a.s.l.)

At 800m (a.s.l.) (figure 5 C3 and C4) the South of CTPVC is represented by medium $\rho$ and high $V_P$ (pink). This region could be related with the formation of Las Cañadas Caldera with high $V_P$ and medium resistivity materials (Coppo et al. 2008).. Around CTPVC, with a ring-shape we can see for the first time the $\rho$ low and $V_P$ low-medium cluster (yellow). This feature could be related with the clay cap of the geothermal system previously identified by Piña-Varas et al. (2014, 2015) or volcanoclastic deposits. The value of the seismic $V_P$ for the central vector of this cluster is 4.5 km/s. At 100 m (a.s.l.) the yellow cluster is still observed around the center of the island. Below the Las Cañadas Caldera a high resistivity and medium $V_P$ (blue colour) is highlight ; it will be a constant feature with increasing depths.

## 4.3 600 m (b.s.l), 1300 m (b.s.l.) and 2000 m (b.s.l.)

At 600 m (b.s.l.) (figure 5 C5, C6 and C7) CTPVC shows high $\rho$ and medium $V_P$ (blue colour), that could be associated to a cold basaltic body. From the center of the island we can see how the resistivity and the $V_P$ decrease towards the coast. The same pattern distribution is present in the depth slices of 1300 m (b.s.l) and 2000 m (b.s.l). Surrounding CTPVC and in the South of the island, some regions show medium ρ and high $V_P$ (pink colour), that could be interpreted as ancient stratovolcano of Roque del Conde (see figure 1 for location). The increase of $V_P$ and ρ values observed below Teide at these depths could be related with the increase of temperature and hence, the bottom of the hydrothermally altered area. This discontinuity between the volcanic edifice and the oceanic crust has been observed by Dañobeitia and Canales (2000) at the same depth. On the other hand, Hernández et al. (2004) related the origin of the fumarolic activity in the top of Teide to a convective gases system in equilibrium. The temperature of the fumarolic activity was hypothesized as if a heat source would be located at sea level or 1 km (b.s.l.). These observations have been already interpreted as geothermal activity in other volcanoes such as Taal volcano in Philippines (Yamaya et al 2013) and Asama, Tarumi and Mt. Fuji volcanoes in Japan (Aizawa et al 2008, Aoki et al 2009, Yamaya et al 2012, Aizawa et al 2005)

### 4.4  2700 m (b.s.l)

In Figure 5 c8 results at 2700 m (b.s.l.) are shown. As can be observed, the distribution of yellow cluster is smaller. On the other hand, the center of the island which is characterized by high resistivity and medium velocity (blue cluster) is surrounded by higher velocity regions (pink cluster). We have interpreted theses results as the presence of a cold basaltic body. García-Yeguas et al (2012) and Prudencio et al (2013) already interpreted the center of Tenerife Island as a cold basaltic body corresponding to the main volcanic edifice based on high velocity and low attenuation results. Finally, at this depth the velocity values decreases towards coastal areas and resistivity results remain at medium values.

## 5   Conclusions

In the present work, we jointly interpreted magnetotelluric and seismic tomography models. We have applied for the first time in volcanic regions FCM technique to obtain 5 clusters based on resistivity and P-wave seismic velocity values. The obtained results allowed us to interpret in more detail the complex structure of Tenerife Island. Based on cluster analysis the most relevant result is the presence of a geothermal system below Teide volcano at 600 m (b.s.l). This interpretation has already made by other authors suggesting that the heat source of the fumarolic activity is located around this depth. Moreover, the clay cap, a typical structure for geothermal systems characterize by low resistivity and medium velocity values and observed in several geothermal environments has been also observed at shallower depths. On the other hand, at deeper depths the results from cluster analysis suggests that the observed structures correspond to ancient volcanic edifices (basaltic bodies) as Roque del Conde in the south. The joint interpretation using FCM has allowed us to identify different interfaces, better constrain the position of the geothermal system and to obtain a more detailed inner structure of Tenerife Island. Although a suitable knowledge of the structure of Tenerife Island is provided with the present study, the combination of new techniques and future development of statistical tools will allow scientific community to reach a higher understanding of volcanic inner structures.

**Acknowledgements**

AGY, JP and JMI   partially granted by the MED-SUV project funded from the European Union´s Seventh Programme for Research, technological development and demonstration under grant agreement No. 308665 and by Grupo de Investigación en Geofísica y Sismología from the Andalusian Regional Program. JP is partially supported by NSF-1066391, NSF-1442630, and NSF- 1125165. JL, AM, PQ and BB partially granted by Project CGL2014-54118-C2-1-R funded by the Spanish Ministry of Economy and Competitiveness, and EU Feder Funds.

**Figure captions**

Fig. 1: Regional setting and location of Tenerife Island. Rift position and Las Cañadas wall are marked with yellow lines. SVZ: Southern Volcanic Zone, PV: Pico Viejo, MB: Montaña Blanca, BTV: Boca Tauce Volcano, SRZ: Santiago Rift Zone, RdC: Roque del Conde and DRZ: Dorsal Rift Zone.

Fig. 2. $F_s$ and $H_s$ coefficients for different number of clusters.

Fig. 3: Cluster distribution for P-wave velocity (km/s) and resistivity (log$\rho$ y log($\Omega$m)). Stars represent the arithmetic average center of each cluster or class. Grey scale indicates the density of P-wave velocity and resistivity values.

Fig. 4. Clusters distribution on Tenerife Island for two cross-sections: North-South (1A-1B) and West-East (2A-2B). Legend: *Blue:* $\rho$ high and $V_P$ (P-wave velocity) medium (RhVm); *Light blue:* $\rho$ medium and $V_P$ low (RmVl); *Yellow:* $\rho$ low and $V_P$ low-medium (RlVlm); *Pink:* $\rho$ medium and $V_P$ high (RmVh); *Green:* $\rho$ medium and $V_P$ medium (RmVm).

Fig. 5: **A1-A8**. Resistivity model images from Piña-Varas et al. (2014).**B1-B8**. Seismic tomography model in absolute P-wave seismic velocity from García-Yeguas et al. (2012). **C1-C8**. Clusters distribution on Tenerife Island. Different rows corresponds to different depths: 2200 m (a.s.l.), 1500 m (a.s.l.), 800 m (a.s.l.), 100 m (a.s.l.), 600 m (b.s.l.), 1300 m (b.s.l.), 2000 m (b.s.l.) and 2700 m (b.s.l.) . Legend: *Blue:* $\rho$ high and $V_P$ (P-wave velocity) medium (RhVm); *Light blue:* $\rho$ medium and $V_P$ low (RmVl);

*Yellow:* ρ low and $V_P$ low-medium (RlVlm); *Pink:* ρ medium and $V_P$ high (RmVh); *Green:* ρ medium and $V_P$ medium (RmVm).

Table 1. In this table are indicated the values of the mean P-wave velocity uncertainties (km/s) and their interval of confidence at different depths.

Table 2. In this table are specified the standard deviation of logarithm resistivity and P-wave velocity variables for every cluster. Moreover, we have added a column describing an interpretation.